\documentstyle[12pt,geompsfi]{article}

\begin{document}
\begin{titlepage}
\begin{flushleft}
Stockholm\\
November 1994\\
\end{flushleft}
\vspace{1cm}
\begin{center}
{\Large NOTE ON MASSIVE SPIN 2 IN CURVED SPACE}\\
\vspace{2cm}
{\large Ingemar Bengtsson}\footnote{Email address: ingemar@vana.physto.se}\\
{\sl Fysikum\\
University of Stockholm\\
Box 6730, S-113 85 Stockholm, Sweden}\\
\vspace{3cm}
{\bf Abstract}\\
\end{center}
We give a Hamiltonian formulation of massive spin 2 in arbitrary Einstein
space-times. We pay particular attention to  Higuchi's forbidden mass range in
deSitter space.
\end{titlepage}

\noindent {\bf 1. INTRODUCTION AND CONCLUSIONS.}

\vspace*{5mm}

\noindent It may be that Nature evades the consistency problems of the higher
spin equations through the simple expedient of not employing them at all. Even
so, and certainly as long as we are not quite sure of Nature's verdict, the
subject remains notoriously fascinating. Here we will study the massive spin 2
model generalized to space-times whose metric obeys Einstein's vacuum
equations, but which are otherwise arbitrary. It is well known that the
Fierz-Pauli equations are algebraically consistent whenever the background
metric has the Einstein property. What is less well known is a peculiar
phenomenon in deSitter space, first studied by Higuchi \cite{Higuchi}
(following an observation originally made by Deser and Nepomechie
\cite{Deser}).

Let us recall what the situation is. In flat space, the Fierz-Pauli equations
are known to describe a field with five degrees of freedom, which carries an
irreducible massive spin 2 representation of the Poincar\'{e}\ group. These
equations remain algebraically consistent in curved space-times whose metric
solves Einstein's vacuum equations. (This does not mean that a single massive
spin 2 field can be coupled to gravity, in fact the consensus is that it can
not \cite{Aragone}. Massive spin 2 fields do appear among the interacting
modes of Kaluza-Klein and string theories, however.) When a positive
cosmological constant is included, it was observed \cite{Deser} that the
equations possess a gauge invariance when $3m^2 = 2{\lambda}$. This was
further elucidated by Higuchi \cite{Higuchi}, who found that the quantum
version of the theory, in deSitter space, has negative norm states whenever
$m^2$ lies below that value. This underlies the further interesting fact
\cite{Higuchi} that there is no van Dam-Veltman discontinuity \cite{Veltman}
in deSitter space. The result can be understood from the point of view of the
representation theory of the deSitter group \cite{Ottoson}. Here we will
understand this phenomenon from yet another, purely classical, point of view,
by verifying explicitly that the Hamiltonian is not positive definite when
$m^2$ belongs to the forbidden mass range.

To be precise, the equations that we will study are

\begin{equation} W_{{\alpha}{\beta}} -
\frac{1}{2}g_{{\alpha}{\beta}}W_{\gamma}^{\ {\gamma}} +
\frac{4{\lambda}}{D-2}h_{{\alpha}{\beta}} -
\frac{2{\lambda}}{D-2}g_{{\alpha}{\beta}}h = m^2(h_{{\alpha}{\beta}} -
g_{{\alpha}{\beta}}h) \ , \label{nr1} \end{equation}

\noindent where $D$ is the dimension of spacetime, $h \equiv h_{\alpha}^{\
{\alpha}}$,

\begin{equation} W_{{\alpha}{\beta}} \equiv
\nabla_{\gamma}\nabla^{\gamma}h_{{\alpha}{\beta}} -
\nabla_{\gamma}\nabla_{\alpha}h_{\beta}^{\ {\gamma}} -
\nabla_{\gamma}\nabla_{\beta}h_{\alpha}^{\ {\gamma}} +
\nabla_{\alpha}\nabla_{\beta}h \ , \end{equation}

\noindent and $\nabla_{\alpha}$ is the covariant derivative compatible with
the the background metric $g_{{\alpha}{\beta}}$, which obeys Einstein's vacuum
equations

\begin{equation} R_{{\alpha}{\beta}} =
\frac{2{\lambda}}{D-2}g_{{\alpha}{\beta}} \ . \end{equation}

\noindent Eq. (\ref{nr1}) implies that $h_{{\alpha}{\beta}}$ is traceless and
obeys the Fierz-Pauli equations

\begin{equation} \nabla_{\gamma}\nabla^{\gamma}h_{{\alpha}{\beta}} +
\frac{4{\lambda}}{D-2}h_{{\alpha}{\beta}} = m^2h_{{\alpha}{\beta}}
\hspace*{25mm} \nabla^{\beta}h_{{\alpha}{\beta}} = 0 \ . \end{equation}

\noindent This is algebraic consistency. It is natural to demand also that the
Hamiltonian should be bounded from below, in suitably chosen backgrounds. This
is the main question to be discussed here.

The above derivation of the Fierz-Pauli equations breaks down for the special
value $(D-1)m^2 = 2{\lambda}$. What happens then \cite{Deser} is that
eq. (\ref{nr1}) is invariant under the gauge transformation

\begin{equation} {\delta}h_{{\alpha}{\beta}} =
\nabla_{\alpha}\nabla_{\beta}{\xi} +
\frac{2{\lambda}}{(D-1)(D-2)}g_{{\alpha}{\beta}}{\xi} \ . \label{Nr5}
\end{equation}

In section 2 we set up the Hamiltonian version of the above equations in
arbitrary Einstein spaces. For flat space, this was first done in
ref. \cite{vanDam}; we have streamlined the derivation through the simple
expedient of relying on results for linearized gravity as far as possible. In
section 3 we study the positivity properties of the Hamiltonian in flat and
deSitter backgrounds. It is shown (not surprisingly, perhaps) that Higuchi's
forbidden mass range can be characterized as that range of mass for which the
Hamiltonian is not bounded from below, thus emphasizing that the phenomenon
has nothing to do with quantum mechanics as such. In section 4 we briefly
consider, and rule out, some deformations of the gauge algebra (\ref{Nr5})
that appears for $3m^2 = 2{\lambda}$. The conclusion of this analysis is that
the simplest attempts to write down a non-linear generalization of this
algebra fail. It is therefore unlikely that there exists an interesting
non-linear version of this somewhat peculiar gauge theory.

The mass of the lightest (unstable) spin 2 meson is 1275 MeV - the square root
of the cosmological constant, in the same units, is known to be very much
smaller indeed, which certainly suggests that Higuchi's forbidden mass range
is of no real interest. Nevertheless we feel that it is worth some attention
as part of the lore of higher spins. As a speculation, we observe that the
exceptional value $3m^2 = 2{\lambda}$ leads to a model which not only has an
extra gauge invariance, but which also exhibits propagation confined to the
light cone \cite{Deser}. Could this serve as a building block for an unknown
long-range force?

\vspace*{1cm}

\noindent {\bf 2. HAMILTONIAN FORMULATION.}

\vspace*{5mm}

\noindent It is not an attractive task to perform a 3+1 decomposition of the
action from which eq. (\ref{nr1}) can be derived. A short cut is to set $m^2 =
0$ at first, in which case all we have to do is to linearize the ADM action
for gravity. The ADM action is

\begin{equation} S_{ADM} = \int \ \dot{{\bf g}}_{ab}{\Pi}^{ab} - {\bf N}{\cal
H} - {\bf N}^a{\cal H}_a \ , \end{equation}

\noindent where

\begin{eqnarray} & {\cal H} = \frac{1}{\sqrt{{\bf g}}}({\bf g}_{ac}
{\bf g}_{bd} - \frac{1}{D - 2}{\bf g}_{ab}{\bf g}_{cd}){\Pi}^{ab}{\Pi}^{cd} -
\sqrt{{\bf g}}(R({\bf g}) - 2{\lambda}) \\
\ \nonumber \\
& {\cal H}_a = - 2 \nabla_b{\Pi}^b_{\ a} \ . \ \end{eqnarray}

\noindent We linearize this around a solution of the equations, according to

\begin{eqnarray} & {\bf g}_{ab} \equiv g_{ab} + h_{ab} \ \ \ \ \ \ \ \
{\Pi}^{ab} \equiv {\pi}^{ab} + p^{ab} \label{nio} \\
\ \nonumber \\
& {\bf N} \equiv N + n \ \ \ \ \ \ \ \ {\bf N}^a \equiv N^a + n^a \
. \label{10} \end{eqnarray}

Note that, from now on, we will use the spatial background metric $g_{ab}$ to
raise and lower spatial vector indices, $h \equiv h_a^{\ a}$, and ${\pi}^{ab}$
is also part of the fixed background. Also, the Ricci tensor and the covariant
derivative that will occur below are defined with respect to the background
metric.

Proceeding in this way, we obtain the following action for linearized gravity:

\begin{equation} S = \int \ \dot{h}_{ab}p^{ab} - N{\cal H}^{(2)} - N^a{\cal
H}^{(2)} - n{\cal H}^{(1)} - n^a{\cal H}^{(1)} \ . \label{lingrav}
\end{equation}

\noindent The constraints of the linearized theory are explicitly

\begin{eqnarray} & {\cal H}^{(1)} = \frac{2}{\sqrt{g}}({\pi}_{ab}p^{ab} -
 \frac{1}{D - 2}{\pi}p + {\pi}^{ac}{\pi}_c^{\ b}h_{ab} - \frac{1}{D - 2}
{\pi}{\pi}^{ab}h_{ab}) - \nonumber \\
\ \\
& \ \ \ \ \ \ \ \ - \sqrt{g}(\nabla_a\nabla_bh_a^{\ b} - \nabla_a\nabla^ah -
R^{ab}h_{ab} + Rh - 2{\lambda}h) \approx 0 \ . \nonumber \\ \ \nonumber \\ &
{\cal H}^{(1)}_a = - 2\nabla_bp^b_{\ a} + {\pi}^{bc}(\nabla_ah_{bc} -
2\nabla_bh_{ca}) \approx 0 \ . \ \ \ \ \ \ \end{eqnarray}

\noindent The Hamiltonian of the linearized theory is quadratic in the fields
$h_{ab}$ and $p^{ab}$. It can be obtained from the ADM phase space action
through a straightforward but lengthy calculation. Unfortunately the explicit
formula that one obtains is also lengthy, and not illuminating to look at, so
we do not give it here. In fact, the entire, explicit formula will not be
needed in the sequel. So we simply write the Hamiltonian of linearized gravity
as

\begin{equation} H \equiv {\cal H}^{(2)}[N] + {\cal H}^{(2)}_a[N^a] \equiv
\int \ N{\cal H}^{(2)} + N^a{\cal H}^{(2)}_a \ . \end{equation}

\noindent Note that $N$ and $N^a$ are part of the fixed background; this
equation also serves to introduce our notation for smearing with test
function.

For all backgrounds that solve Einstein's equations, the linearized constraint
algebra - which is abelian - can be found through linearization of the
constraint algebra of the non-linear theory. Moreover, by studying the second
order terms of the non-linear constraint algebra one can deduce that

\begin{eqnarray} & \frac{d}{dt}{\cal H}^{(1)}[n] = \frac{\partial}{dt}
{\cal H}^{(1)}[n] + \{{\cal H}^{(1)}[n], H\} = \ \ \ \ \ \ \ \nonumber \\
\  \label{tidsder} \\
& \ \ \ \ \ = {\cal H}_a^{(1)}[n\nabla^aN - N\nabla^an] - {\cal
H}^{(1)}[N^a\nabla_an] \approx 0 \ .  \nonumber \end{eqnarray}

\noindent Hence the linearized version of the Hamiltonian constraint is
conserved under time evolution, as is indeed required by the consistency of
linearized gravity. Eq. (\ref{tidsder}) - which would have been awkward to
prove through directly, using an explicit form of the Hamiltonian - will be
used below.

With linearized gravity in hand, the Hamiltonian formulation of the massive
theory can be derived in a few steps. First we add a Pauli-Fierz mass term to
the action:

\begin{equation} S_m \equiv \int
\sqrt{-g}(h_{{\alpha}{\beta}}h^{{\alpha}{\beta}} - hh) \ . \end{equation}

\noindent Now, by comparing to the ADM decomposition of the space-time metric,
it is easy to see that eqs. (\ref{nio} - \ref{10}) implies the following
``names'' for the time-time and time-space components of the field
$h_{{\alpha}{\beta}}$ that occur here:

\begin{eqnarray} & h_{ta} = n_a + N^bh_{ba} \\
\ \nonumber \\
& h_{tt}= - 2nN + 2n_aN^a + N^ah_{ab}N^b \ . \end{eqnarray}

It is then a matter of straightforward calculation to show that

\begin{equation} \sqrt{-g}(h_{{\alpha}{\beta}}h^{{\alpha}{\beta}} - hh) =
N\sqrt{g}(h_{ab}h^{ab} - h^2) - \frac{2\sqrt{g}}{N}n_an^a - 4\sqrt{g}nh \
. \end{equation}

\noindent In this form, these terms are to be added to the linearized gravity
phase space action (\ref{lingrav}). We observe that the Lagrange multiplier
$n^a$ then occurs quadratically, so that it can be solved for. Inserting the
result back into the action, we are then left with

\begin{eqnarray} & S = \int \ \dot{h}_{ab}p^{ab} - N{\cal H}^{(2)} -
 \frac{m^2N}{4}\sqrt{g}(h_{ab}h^{ab} - h^2) - \ \ \ \ \nonumber \\
\ \\
& \ \ \ \ - \frac{N}{2m^2\sqrt{g}}{\cal H}_a^{(1)}{\cal H}^{(1)a} - n({\cal
H}^{(1)} - m^2\sqrt{g}h) \ . \nonumber \end{eqnarray}

\noindent If any other form of the mass term than that given by Fierz and
Pauli is being used one finds that also $n$ occurs quadratically here. When
this Lagrange multiplier is eliminated from the action one finds a model with
no constraints on the six degrees of freedom. The ``extra'' degree of freedom
is a negative energy ghost in Minkowski space (where such a statement makes
sense).

We must now compute the secondary constraints, if any, by taking the total
time derivative of the constraint that is obtained when the action is varied
with respect to the Lagrange multiplier $n$. It is here that
eq. (\ref{tidsder}) comes in handy. After a non-trivial but not too lengthy
calculation, one finds in this way that the degrees of freedom are subject to
the following constraints:

\begin{eqnarray} & {\Psi}_1 = {\cal H}^{(1)} - m^2\sqrt{g}h \approx 0 \\
\ \nonumber \\
& {\Psi}_2 = - \frac{1}{2}\nabla^a{\cal H}_a^{(1)} + \frac{m^2}{D - 2}p -
m^2\frac{D - 4}{D - 2}{\pi}^{ab}h_{ab} \approx 0 \ . \end{eqnarray}

\noindent The algebra of these constraints is

\begin{eqnarray} \{ {\Psi}_1(x), {\Psi}_2(x')\} = - m^2(m^2 -
\frac{2{\lambda}}{D - 1})\frac{D - 1}{D - 2} \sqrt{g}{\delta}(x,x') \
. \end{eqnarray}

\noindent We see that the constraints are second class whenever

\begin{equation} (D - 1)m^2 \neq 2{\lambda} \ . \end{equation}

\noindent For all such values of $m^2$ then, the model possesses the expected
five degrees of freedom. The special value of $m^2$ for which the constraint
algebra is first class (and the number of degrees of freedom drops to four) is
precisely the value found in ref. \cite{Deser} - as had to be the case, of
course.

\vspace*{1cm}

\noindent {\bf 3. POSITIVITY PROPERTIES.}

\vspace*{5mm}

\noindent For suitable background space-times, we expect the Hamiltonian to be
bounded from below. To show this one will have to employ a decomposition of
the field variable into irreducible parts, which in general is not
straightforward. Therefore we confine our attention to deSitter space, and use
the coordinate system

\begin{equation} ds^2 = - dt^2 + e^{2{\alpha}t}(dx^2 + dy^2 + dz^2)
 \ , \end{equation}

\noindent where

\begin{equation} (D - 1){\alpha}^2 = \frac{2{\lambda}}{D - 2} \
.\end{equation}

\noindent Then we can use flat space projection operators to perform the
decomposition, viz.

\begin{eqnarray} & h^{(2)}_{ab} = ({\Pi}_a^{\ (c}{\Pi}_b^{\ d)} -
\frac{1}{3}{\Pi}_{ab}{\Pi}^{cd})h_{cd} \ \ \ \ \ \nonumber  h^{(0)}_{ab}
 = \frac{1}{3}{\Pi}_{ab}{\Pi}^{cd}h_{cd} \\
\ \\
& h^{(1)}_{ab} = ({\Pi}_a^{(c}{\Omega}_b^{d)} + {\Omega}_a^{\ (c}{\Pi}_b^{\
d)})h_{cd} \ \ \ \ \ \ h^{(0')}_{ab} = {\Omega}_{ab}{\Omega}^{cd}h_{cd} \ ,
\nonumber \end{eqnarray}

\noindent where

\begin{equation} {\Pi}_a^{\ b} = {\delta}_a^{\ b} - \frac{1}{\triangle
}\partial^a\partial^b \ \ \ \ \ \ \ \ \ \ \ \ \ {\Omega}_a^{\ b} =
\frac{1}{\triangle }\partial_a\partial^b \ , \end{equation}

\noindent ${\triangle}$ is the Laplacian, and the spatial dimension was set to
three for simplicity ($D = 4$). Actually there is a problem here, because in
deSitter space the fall-off behaviour of the fields is not that which is
required to justify our manipulations with the projection operators. However,
we can always restrict our attention to initial data with compact
support. This will be enough to show that the Hamiltonian is not bounded from
below under certain conditions, although strictly speaking it is not enough to
show that it is bounded from below under other conditions.

When we use the chosen form of the deSitter metric as our background metric in
the Hamiltonian, we find - although the reader will have to take this on
trust, since we did not give the explicit form of the Hamiltonian in an
arbitrary background - that

\begin{eqnarray} & H = \int \frac{1}{\sqrt{g}}(p^{ab}p_{ab} - \frac{1}{2}pp)
 + \nonumber \\
\ \nonumber \\
& \ \ + \frac{\sqrt{g}}{4}(h\triangle h - h_{ab}\triangle h_{ab} -
2\partial_bh^{ab}\partial^ch_{ac} + 2\partial^ah\partial^bh_{ab}) + \nonumber
\\ \ \\ & + {\alpha}(hp - 2h_{ab}p^{ab}) + {\alpha}^2\sqrt{g}h_{ab}h^{ab} +
\nonumber \\ \ \nonumber \\ & + \frac{m^2}{4}\sqrt{g}(h_{ab}h^{ab} - h^2) +
\frac{2}{m^2\sqrt{g}}\partial_bp^{ab}\partial^cp_{ac} \ . \nonumber
\end{eqnarray}

\noindent The spatial background metric (which has been used to raise and
lower indices) is flat but time dependent, so this is a time dependent
Hamiltonian.

The constraints affect only the scalar modes:

\begin{eqnarray} & {\Psi}_1 = \sqrt{g}(\triangle h^{(0)} + (2{\alpha}^2 -
 m^2)h) + 2{\alpha}p = 0 \\
\ \nonumber \\
& {\Psi}_2 = 2\triangle p^{(0')} + m^2p + 2{\alpha}\sqrt{g}(2\triangle
h^{(0)}
- \triangle h) = 0 \ . \end{eqnarray}

\noindent (Recall that $h \equiv h_a^{\ a}, h^{(0)} \equiv h_{\ a}^{(0)\ a}$
etc.) We now have to decide on a set of independent data, in terms of which
the constraints are to be solved for the remaining data. It turns out that a
convenient set of independent data is $h^{(0)}$ and

\begin{equation} p' \equiv p^{(0')} + {\alpha}\sqrt{g}(2h^{(0)} - h) \
. \end{equation}

We now turn to the positivity properties of the Hamiltonian. In a flat
spacetime it is not too difficult to show, using a decomposition of the fields
into irreducible parts, that the Hamiltonain is positive semi-definite. In
deSitter space on the other hand - as the reader may verify - an attempt to
express the Hamiltonian as a function

\begin{equation} H = H(h^{(2)}, p^{(2)}, h^{(1)}, p^{(1)}, h^{(0)}, p')
\end{equation}

\noindent leads to a surprisingly lengthy calculation. Therefore we simplify
matters further by choosing initial data such that $p' = 0$. Then one can show
that

\begin{eqnarray} & H_{\mid p'=0} = \nonumber \\
\ \nonumber \\
& = \int \frac{1}{\sqrt{g}}(p^{(2)}_{ab} -
{\alpha}\sqrt{g}h^{(2)}_{ab})(p^{(2)ab} - {\alpha}\sqrt{g}h^{(2)ab}) +
\frac{\sqrt{g}}{4}\partial_ch^{(2)}_{ab}\partial^ch^{(2)ab} +
\frac{m^2\sqrt{g}}{4}h^{(2)}_{ab}h^{(2)ab} + \nonumber \\ \ \nonumber \\ & +
\frac{1}{\sqrt{g}}(p^{(1)}_{ab} - {\alpha}\sqrt{g}h^{(1)}_{ab})(p^{(1)ab} -
{\alpha}\sqrt{g}h^{(1)ab}) +
\frac{1}{m^2\sqrt{g}}\partial_cp^{(1)}_{ab}\partial^cp^{(1)ab} +
\frac{m^2\sqrt{g}}{4}h^{(1)}_{ab}h^{(1)ab} + \nonumber \\ \ \nonumber \\ & +
\sqrt{g}(\frac{3m^2 + 2{\alpha}^2}{8(m^2 -
2{\alpha}^2)}\partial_ah^{(0)}\partial^ah^{(0)} +
\frac{3m^2}{8}h^{(0)}h^{(0)}) + \\ \ \nonumber \\ & +
{\alpha}^2\sqrt{g}(\frac{1}{2}(h - h^{(0)})^2 + h^{(0)}h^{(0)} +
\frac{2}{m^2}\partial_a(h - 2h^{(0)})\partial^a(h - 2h^{(0)}) + \nonumber \\
 \
\nonumber \\ & + 4\partial_ah^{(0)}\partial^ah^{(0)}) \ . \nonumber
\end{eqnarray}

\noindent By inspection of this expression, we see that there is a term among
the scalar modes which can assume negative values if $3m^2 < 6{\alpha}^2 =
2{\lambda}$, which proves that the Hamiltonian can not be positive
semi-definite for such values of $m^2$.

\vspace*{1cm}

\noindent {\bf 4. ATTEMPTS TO DEFORM THE GAUGE ALGEBRA.}

\vspace*{5mm}

\noindent Before we end this note, we will briefly consider whether it is
possible to deform the algebra of the peculiar gauge symmetry which arises for
$3m^2 = 2{\lambda}$. The question we ask is whether there exists a non-linear
theory invariant under a non-abelian gauge group for which, in a linearized
limit, the abelian gauge transformation (\ref{Nr5}) is recovered. This kind of
question has been asked in the past concerning higher spin gauge algebras, and
it is known to be very difficult to deform any abelian gauge algebra involving
more than one derivative in such a manner \cite{Anders}. (More recently, there
has been progress in higher spin theory, but we will not go into this here.
See ref. \cite{Vasiliev} and references therein.)

So we search for a non-linear transformation of the (somewhat schematically
written) form

\begin{equation} {\delta}h_{{\alpha}{\beta}} =
\nabla_{\alpha}\nabla_{\beta}{\xi} + g_{{\alpha}{\beta}}{\xi} +
(\nabla^2{\xi}h)_{{\alpha}{\beta}} \ . \label{deform} \end{equation}

\noindent Now one can write down a general Ansatz for the structure constants,
and check whether it can obey the Jacobi identity \cite{Anders}. Note that
this kind of argument is very general; it is capable of ruling out also
transformations much more general than that sketched here, such as
transformations that involve several tensor fields in a non-diagonal fashion.

As long as the field $h_{{\alpha}{\beta}}$ carries no internal index, the only
possible form that the commutator of two non-linear transformations can take
is, to lowest order,

\begin{equation} [{\delta}_{\xi}, {\delta}_{\eta}]^{(0)} =
{\delta}_{{\eta}\nabla^2{\xi} - {\xi}\nabla^2{\eta}} \ . \label{commut}
\end{equation}

\noindent However, it is then easy to check that this commutator does not obey
the Jacobi identity

\begin{equation} [{\delta}_{\omega}, [{\delta}_{\xi}, {\delta}_{\eta}]] +
[{\delta}_{\eta}, [{\delta}_{\omega}, {\delta}_{\xi}]] + [{\delta}_{\xi},
[{\delta}_{\eta}, {\delta}_{\omega}]] = 0 \ . \end{equation}

\noindent It follows that there does not exist any deformation of the abelian
gauge algebra into a non-abelian algebra of the suggested type.

A little bit more generally, one may introduce an internal symmetry index $i$,
which permits the Ansatz

\begin{equation} a_1f_{ijk}\nabla_{\gamma}{\xi}^j\nabla^{\gamma}{\eta}^k +
a_2f_{ijk}(\Box {\xi}^j{\eta}^k + {\xi}^j\Box {\eta}^k) +
a_3f_{ijk}{\xi}^j{\eta}^k \end{equation}

\noindent for the zeroth order structure constants in
eq. (\ref{commut}). Going through the Jacobi identity again, with the
$f_{ijk}$'s obeying the Jacobi identity for an ordinatry Lie algebra, one
finds that $a_1 = a_2 = 0$, so this seems to be of no interest either.

We conclude that that it is unlikely that any deformation of the gauge algebra
under study exists.

\end{document}